\documentclass[final]{svjour3}
\usepackage{graphicx}
\usepackage{rotating}
\usepackage{amssymb}
\usepackage{mathptmx}
\usepackage{textcomp}
\usepackage[numbers]{natbib}
\makeatletter
\journalname{Journal of Low Temperature Physics}

\bibpunct{}{}{,}{s}{}{,}

\begin{document}

\newcommand{\hdblarrow}{H\makebox[0.9ex][l]{$\downdownarrows$}-}
\title{Flat low-loss silicon gradient index lens for millimeter and submillimeter wavelengths}

\author{F. Defrance \and C. Jung-Kubiak \and S. Rahiminejad \and T. Macioce \and J. Sayers \and J. Connors \and S. Radford \and G. Chattopadhyay \and S. Golwala}

\institute{Department of Astronomy, California Institute of Technology,\\ Pasadena, CA 91125, USA\\
\email{fdefranc@caltech.edu}}

\maketitle

\begin{abstract}

We present the design, simulation, and planned fabrication process of a flat high resistivity silicon gradient index (GRIN) lens for millimeter and submillimeter wavelengths with very low absorption losses.  The gradient index is created by subwavelength holes whose size increases with the radius of the lens.  The effective refractive index created by the subwavelength holes is constant over a very wide bandwidth, allowing the fabrication of achromatic lenses up to submillimeter wavelengths.  The designed GRIN lens was successfully simulated and shows an expected efficiency better than that of a classic silicon plano-concave spherical lens with approximately the same thickness and focal length.  Deep reactive ion etching (DRIE) and wafer-bonding of several patterned wafers will be used to realise our first GRIN lens prototype.


\end{abstract}

\section{Introduction}

\begin{figure}[htbp]
	\begin{center}
	   \includegraphics[width=6.5cm]{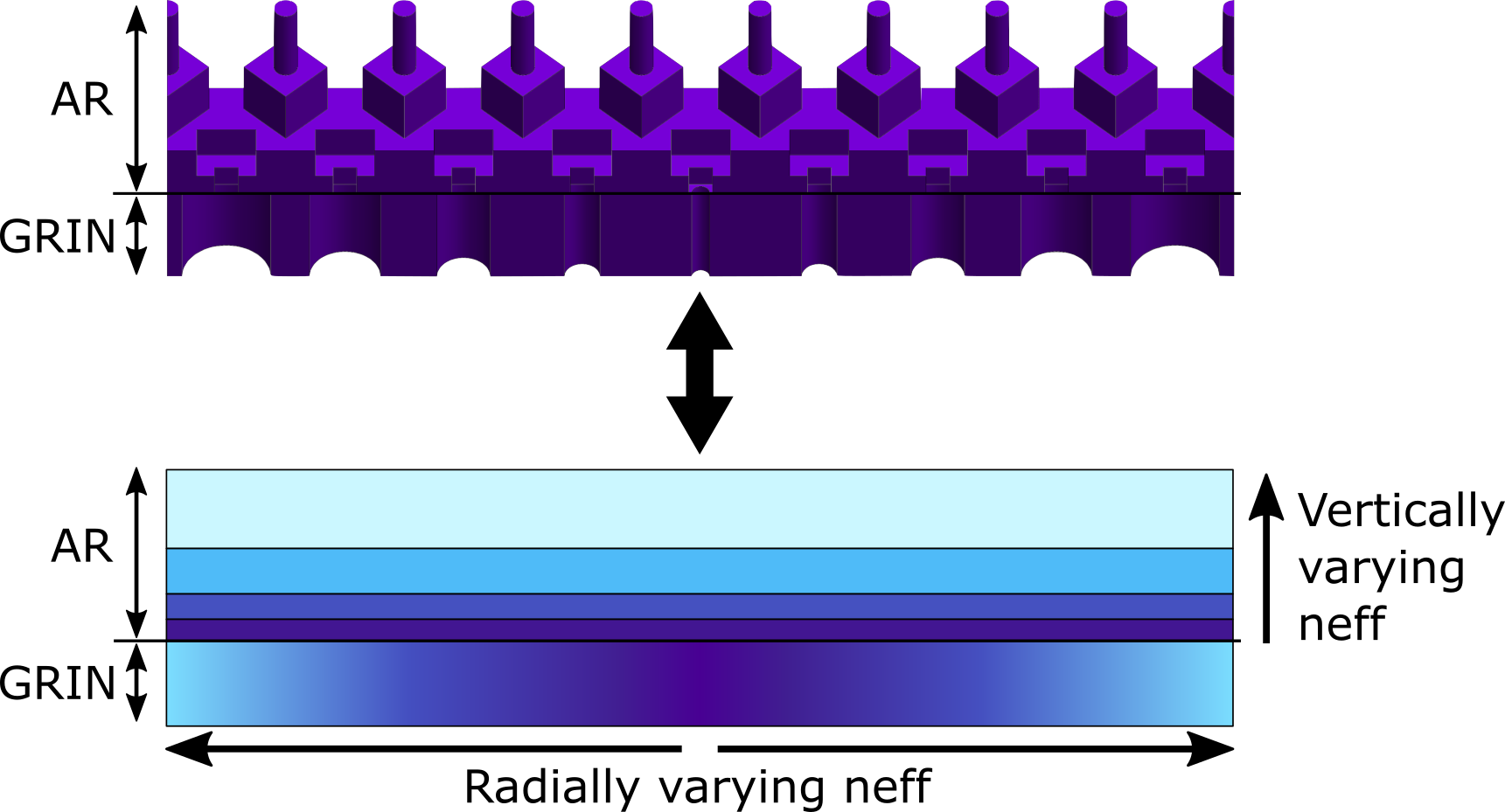}
        \caption{(color online) Schematic of a GRIN lens with a 4-layer AR coating: Effective refractive index is varied vertically for AR layers and radially for the GRIN lens.}
        \label{fig:GRIN+AR}
    \end{center}
\end{figure}

Many applications in astronomy from tens of GHz to THz frequencies, such as CMB polarization studies and Sunyaev-Zeldovich effect observations, would benefit from low loss and wide bandwidth optics.  High resistivity silicon (HRSi) is an excellent material for optics within this frequency range because of its high refractive index ($n_{Si} = 3.42$~\cite{Grischkowsky:90, Dai:04}), achromaticity, lack of birefringence, low loss~\cite{Parshin:1995}, high thermal conductivity, and strength.  Silicon's high index, however, presents a challenge for antireflection (AR) treatment, which our approach addresses.  
To develop wide bandwidth and low loss silicon optics, we are focusing on two key elements: 1) the fabrication of multilayer AR structures via multi-depth deep reactive ion etching (DRIE) and wafer-bonding; and 2) the assembly of gradient index (GRIN) optics, flat-faced to be consistent with AR treatment, by bonding multiple silicon wafers patterned with the desired radial index profile by DRIE.  
The fabrication of 1-layer~\cite{Gallardo:17,Wheeler:14,Wada:2010,Wagner-Gentner:06,Schuster:05} and 2-layer~\cite{Gallardo:17, Defrance:18} AR structures for THz frequencies has already been demonstrated, such as wafer bonding of unpatterned~\cite{Tong:99,Gosele:98,Suni:06,Gallardo:17} and patterned~\cite{Makitsubo:17} silicon wafers.  
Prior attempts to fabricate GRIN lenses for millimeter and submillimeter wavelengths, using silicon and DRIE~\cite{Tang:2015,Park:2014,Brincker:2016}, or other materials~\cite{Moseley:2019}, are extremely promising but do not meet yet the demanding loss, reflectance, and bandwidth requirements on optics for CMB polarization studies and space missions.   
To this end, we are currently developing a new 6-layer AR design~\cite{Macioce:2019} which will provide less than 1\% reflectance over a 6:1 bandwidth, and a GRIN lens model, presented in this article, which should present less than 1\% power absorption losses and stay achromatic over a 6:1 bandwidth.
Both the AR structures and the GRIN structures are made of subwavelength features (posts or holes), etched with DRIE, that change the effective refractive index of silicon.  
In GRIN lenses, the refractive index varies radially (higher in the middle and lower near the edge, for a focusing lens), and the subwavelength structures used need to be holes (no posts) so the lens can be physically continuous and thus edge-mountable.  
To reach any desired GRIN lens thickness, several identical etched wafers must be bonded together because we cannot use DRIE to etch vertical holes deeper than a few hundreds of $\mu$m with a high aspect ratio (up to approximately 20:1).  Once the GRIN lens will be developed, we will be able to combine it with our multilayer AR structures in order to obtain a very low loss, wide bandwidth, GRIN lens for millimeter and submillimeter wavelengths, as shown in fig.~\ref{fig:GRIN+AR}.
We present our results to date, which include the design, simulation, and fabrication method of a 80~mm diameter flat GRIN lens made of high resistivity silicon.

\section{GRIN lens design}

\subsection{Hole and grid geometry}

To design our GRIN lens, we considered square, hexagonal, and round hole geometries.
Hexagonal holes present the advantage of having 120\textdegree~angles which are easier to etch with DRIE than square holes with sharper 90\textdegree~angles which can more easily get rounded during etching.  The maximum aspect ratio achievable with DRIE determines the size of the smallest holes that can be etched.  For square holes, the "lateral" aspect ratio (considering the distance between two parallel faces), which is 1.41 times higher than the "diagonal" aspect ratio (considering the distance between two opposite corners), will limit the minimum size of the holes and the highest effective index that can be reached.  For hexagonal holes, the ratio between the "lateral" and "diagonal" aspect ratios is only 1.15, so for the same "lateral" aspect ratio as for square holes, it is possible to have smaller holes and reach a higher effective index.  Round holes, on the contrary, are the best considering etch angle and aspect ratio, but the maximum size they can reach before touching each other is smaller.  
Therefore, we have chosen to design our GRIN lens with hexagonal holes distributed along an hexagonal grid as it appears to be the best compromise for small and big holes.  A slightly better solution would be to use round geometry for small holes to optimize the aspect ratio, then switch to hexagonal holes when the size increases.  But, this complication did not seem necessary for our first prototype.

\subsection{Effective index}

\begin{figure}[htbp]
	\begin{center}
	    \includegraphics[width=8cm]{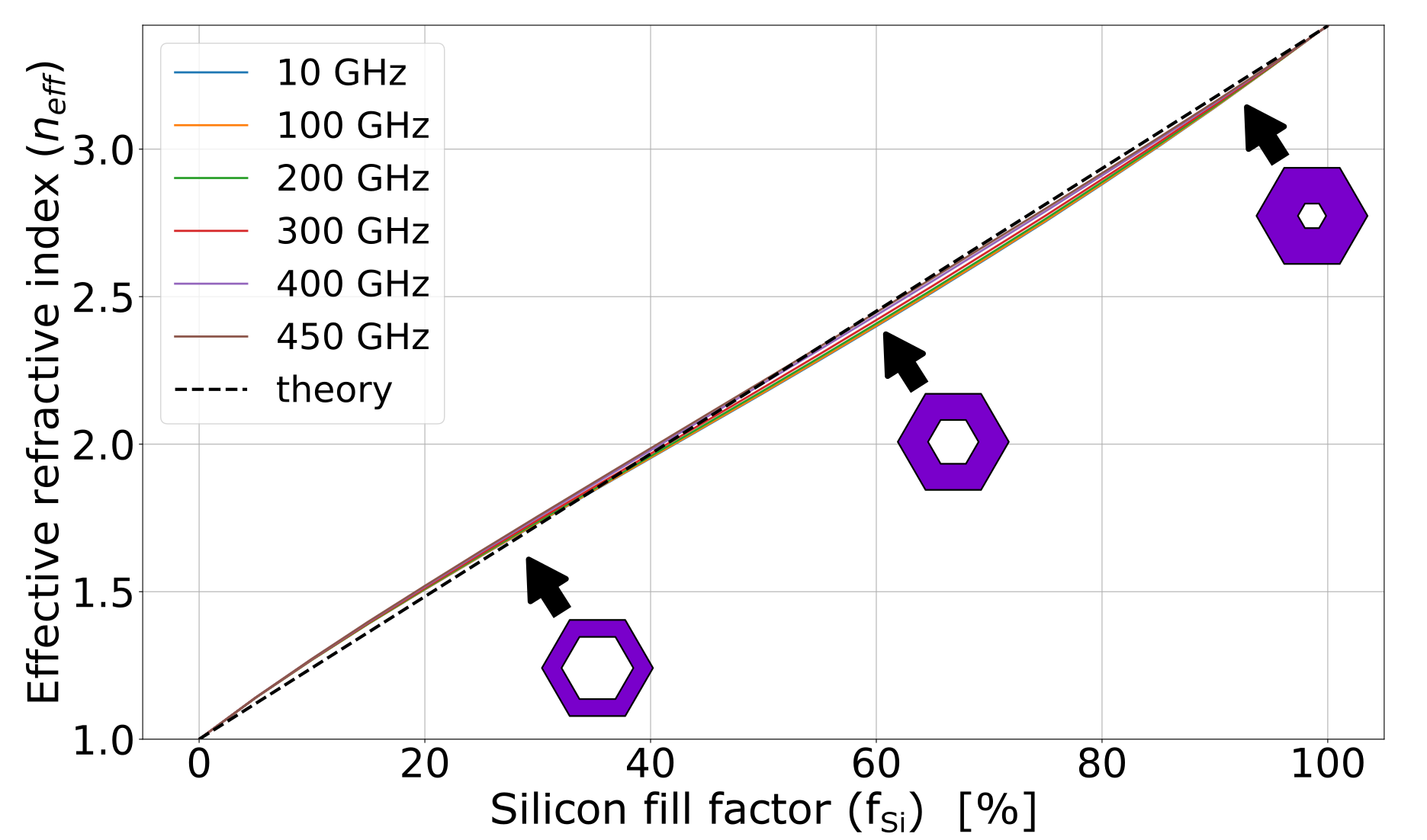}
        \caption{(color online) Plot showing the effective index variation of sub-wavelength hexagonal structures, for different frequencies and fill factors.} 
        \label{fig:HFSS_neff}
	\end{center}
\end{figure}

To design GRIN optics with subwavelength structures, we need to know how the effective refractive index ($n_{eff}$) of silicon varies with the size of holes and frequency.  For this study, we used a commercial electromagnetic finite element solver, ANSYS High Frequency Structure Simulator (HFSS) to calculate the complex reflection and transmission spectra (S parameters) of identical hexagonal holes, distributed over an infinite silicon substrate.  We then fit the spectra with a dielectric slab model to extract the corresponding effective refractive index ($n_{eff}$) as a function of fill factor ($f_{Si} = 1-A_h/A_c$) and frequency, with $A_h$ and $A_c$ the areas of the hole and considered cell, respectively.  
To avoid grating lobes and diffraction effects, the grid spacing, $\Lambda$, must be considerably smaller than the vacuum wavelength, $\lambda/\Lambda > ( n_{Si} + \cos \theta ) \approx 4$, for incident angles $\theta \le 35$\textdegree~\cite{Morris:1993ir}.  We want our GRIN lens to be able to operate up to 420~GHz, so the grid spacing must be smaller than 178~$\mu$m.  However, even within this limit, our simulations show that the variation of the effective index increases when the grid spacing gets closer to the wavelength in the dielectric.  As we also require our GRIN lens to stay achromatic over a wide frequency range ([70, 420]~GHz),  we chose a grid spacing of 75~$\mu$m which is a good compromise between $n_{eff}$ steadiness over bandwidth and fabrication constraints.  
As shown in fig.~\ref{fig:HFSS_neff}, for 75~$\mu$m grid spacing, the index variation up to 450~GHz is negligible.  
The variation of $n_{eff}$ with the fill factor is usually considered linear but our simulations show a small deviation from this linear relation.  We fitted the results of these simulations with a polynomial to determine the hole size needed to generate any desired $n_{eff}$.
We also performed additional simulations and confirmed that the variation of the polarisation ($\theta$) of the TE incident plane wave did not have any impact on $n_{eff}$ (with $-20$\textdegree $< \theta < 20$\textdegree). 

\subsection{Dimensions, characteristics and profile}

We designed a 80~mm diameter GRIN lens, with 75~mm focal length.  According to previous DRIE tests, it has been proven difficult to accurately etch small holes with an aspect ratio higher than 15:1 (the wall of the etched holes is not very vertical when the aspect ratio is too high).  As we are using 500~$\mu$m thick silicon wafers, and the holes are etched from both sides of wafers, the etching depth is 250~$\mu$m, which limits our holes' diameter to approximately 17~$\mu$m (for safety we chose a minimum diameter of 19~$\mu$m).  The maximum size of the holes is limited by the minimum thickness of silicon that needs to separate the holes to preserve the strength of the lens.  We have estimated this minimum "wall" thickness to approximately 15~$\mu$m.  With a grid spacing of 75~$\mu$m, it gives a maximum hole diameter of 60~$\mu$m.  By using the data shown in fig.~\ref{fig:HFSS_neff}, we deduced that these minimum and maximum hole dimensions correspond to effective indexes: $n_{min} = 1.87$ and $n_{max} = 3.25$. 
The variation of the index along our GRIN lens radius follows a parabolic profile: 

\begin{equation}
\label{eq1}
    n(r) = n(0) - \frac{r^2}{2 f t},
\end{equation}{}

with r the radial distance to the center of the lens, $n(0) = n_{max}$ the index at $r = 0$, $f = 75~mm$ the focal length, and $t$ the thickness of the lens.  As the radius of the lens is $40~mm$, we must have $n(40~mm) = n_{min}$, and we can easily deduce the required thickness of the GRIN lens: $t \approx 7.7~mm$.

\section{GRIN lens simulation}

\begin{figure}[htbp]
	\begin{center}
	    \includegraphics[width=5.5cm]{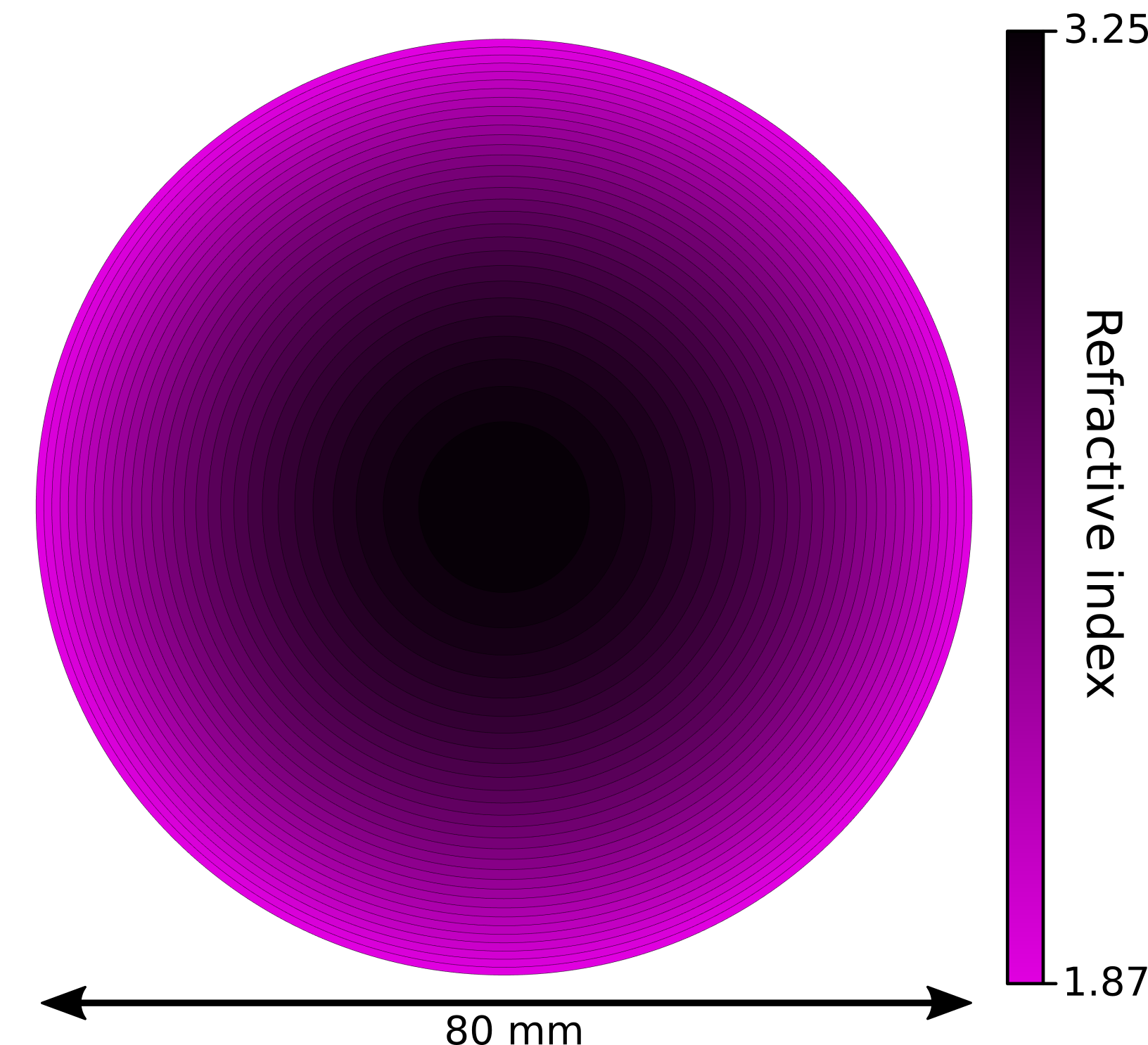}
        \caption{(color online) GRIN lens model for Feko with 30 discrete annuli of different indexes} 
        \label{fig:feko_GRIN_model}
	\end{center}
\end{figure}

Due to the large size of the GRIN lens, compared to the wavelength, it would be computationally very demanding to simulate its real structure with hundreds of thousands of hexagonal holes.  Therefore, with eq.~(\ref{eq1}), we designed a lens model made of 30 concentric annuli with 30 different indexes linearly spaced between $n_{min}$ and $n_{max}$, as shown in fig.~\ref{fig:feko_GRIN_model}.  The number of annuli was increased up to 50 without showing any change in the simulation results, which indicates that 30 annuli is enough to simulate the GRIN lens.

\begin{figure}[htbp]
	\begin{center}
	    \includegraphics[width=12cm]{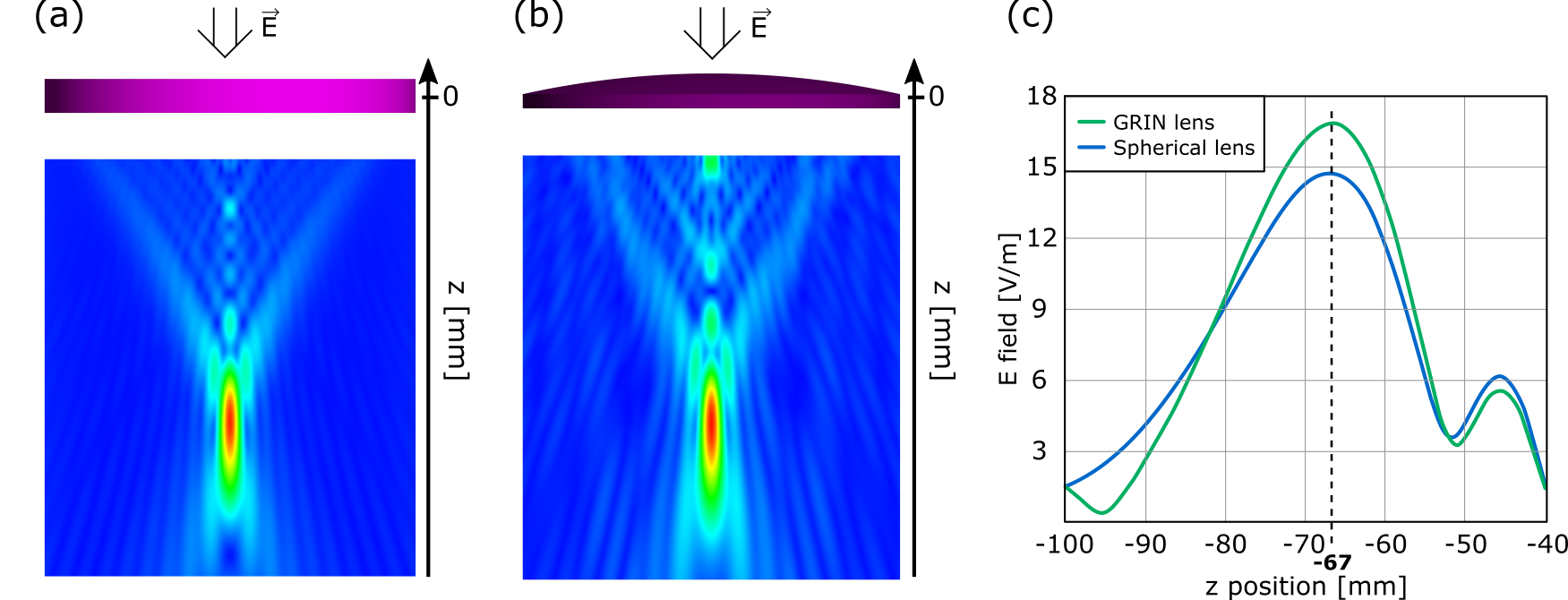}
        \caption{(color online) Electric field amplitude distribution, simulated with Feko, behind (a) the GRIN lens model made of 30 annuli with different indexes and (b) the silicon plano-concave spherical lens model. The comparison (c) of the electric field amplitude along the optical axis for both lenses shows that they have the same focal length and the GRIN lens has a higher focusing efficiency.} 
        \label{fig:feko_nf_result}
	\end{center}
\end{figure}

Each annulus is made of plain homogeneous material, making the model much easier to simulate.  For the simulations, we used the commercial software Feko with the solver RL-GO (Ray Launching Geometrical Optics).  
A 100~GHz plane wave with TE (transverse electric) polarization was used to illuminate the lens, and the magnitude of the electric field was calculated behind the lens, along the optical axis.  Two simulations were performed, with the GRIN lens, and with a classic silicon plano-concave spherical lens with the same diameter and focal length.  The results of both simulations are shown in fig.~\ref{fig:feko_nf_result}.  We see that the electric field distribution is very similar for both lenses, and they have the same focal length, 67~mm, which is slightly shorter than the designed one.  The GRIN lens has a higher focusing efficiency than the plano-concave spherical lens, with 25\% more power at the focal point, probably because the GRIN lens does not have any spherical aberration and has less reflection at the interface with air due to its lower index.  

\section{Fabrication method}

\begin{figure}[htbp]
	\begin{center}
	    \includegraphics[width=9cm]{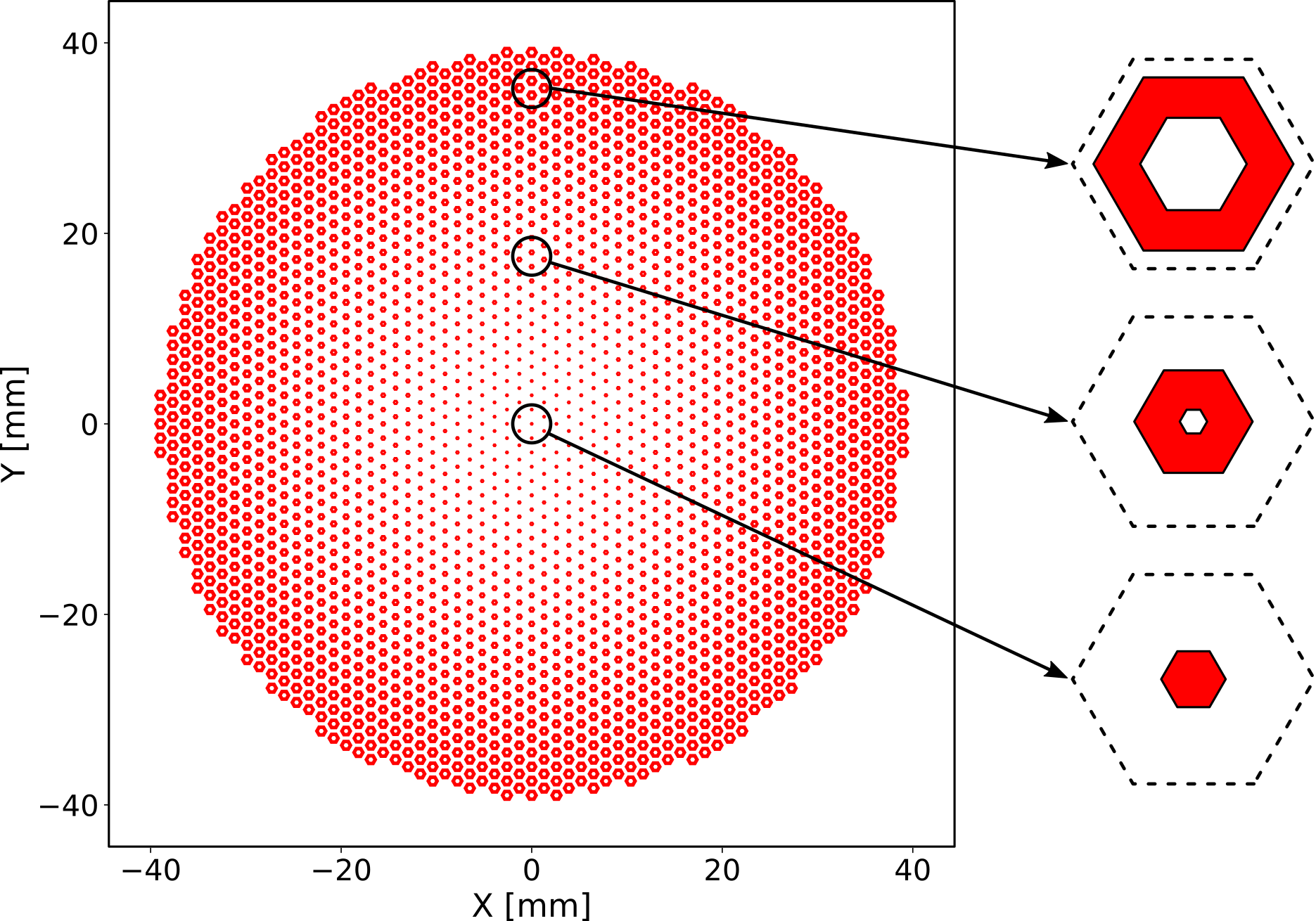}
        \caption{(color online) Schematic of the new mask design (with bigger cells so they can be distinguished on the drawing), with zoomed cells to distinguish the new hole pattern for similar etching speed.} 
        \label{fig:mask_with_posts}
    \end{center}
\end{figure} 

The GRIN lens will be made with 100~mm diameter, 500~$\mu$m thick, high resistivity silicon wafers, stacked and bonded together.  To achieve the focal length of 75~mm of the designed GRIN lens, we will need a total thickness of approximately 7.7~mm, which requires the bonding of 15 wafers.  DRIE is used to etch each wafer\cite{Jung-Kubiak:16}, from both sides (so we will only need to etch 250~$\mu$m deep holes).  We have already successfully achieved DRIE patterning and bonding of silicon wafers to create an antireflection treatment with subwavelength structures\cite{Defrance:18}.  We are applying the same technique to fabricate the GRIN lens.  A first etch test, however, showed that the small holes (in the middle of the lens) did not etch through, while the biggest holes (near the edge of the lens) over-etched.  To solve this problem, we need all holes to be etched at a similar speed.  As the etching speed depends on the aspect ratio of the holes, we invented a new method to keep a similar aspect ratio for all holes.  
As shown in fig.~\ref{fig:mask_with_posts}, for big holes, we only etch the outer annulus of the hole and keep a pillar in the middle.  This method will slow down the etching speed of the bigger holes, and when the annulus will be etched completely through the wafer, the middle pillar will fall, creating the intended big hole.  The mask corresponding to this new design has been created and the fabrication test will happen in a very near future.

\section{Conclusion}

We have successfully designed and simulated a gradient index lens for millimeter and submillimeter wavelengths.  The effective refractive index of hexagonal holes was studied and simulated to obtain a precise relation between the fill factor, the frequency and the effective index.  This study was then used to build the GRIN lens profile.  The simulation results of the GRIN lens are very similar to those obtained with a classic plano-convex spherical lens, and therefore they seem reasonable.  The efficiency of the simulated GRIN lens is already higher than that of the spherical lens, and the addition of antireflection structures will even increase the efficiency.  Finally we come with an innovative solution to etch different hole sizes at a similar speed with DRIE, and the fabrication of our first GRIN lens planned in a very near future.

\begin{acknowledgements}
This work was funded by NASA (Grant NNX15AE01G).  The Caltech/JPL President’s and Director’s Research Fund proposal was recently successful and will allow us to continue this work.  SBIR and APRA proposals are currently under review.
\end{acknowledgements}

\bibliographystyle{ieeetr}
\bibliography{bibliography}

\pagebreak

\end{document}


\newcommand{\hdblarrow}{H\makebox[0.9ex][l]{$\downdownarrows$}-}
\title{Title}

\author{A.B. Name \and C.D. Name}

\institute{Department of Physics, Name University,\\ City, STATE zip, Country\\ Tel.:\\ Fax:\\
\email{name@email.com}}

\maketitle

\begin{abstract}

Insert abstract text directly after the heading. Leave an empty line after the abstract. Abstract should be one paragraph, with no indentation, and justified.

\keywords{keyword 1, keyword 2...}

\end{abstract}

\section{Section Title}

When preparing your manuscript, please follow the instructions in this template. This template serves as a universal template for Special Issue Articles to be taken into consideration for publication in the Journal of Low Temperature Physics.\\
Insert an empty line after the section title. Indent at the start of each paragraph after the first paragraph of the section, which is not indented. For special issue articles the page length is 6 pages. Invited contributions may be 12 pages long. In order to estimate your article length, please prepare your manuscript in Times or Times New Roman, font size 11, justify the body text, and make sure the page format is set to A4. The template margins are: Top: 2.07”, Left: 1.77”, Bottom: 2.17”, Right: 1.77” inches.\\
Subsections can be used. 

\section{Section Title}

\begin{figure}[htbp]
\begin{center}
\includegraphics[width=0.8\linewidth, keepaspectratio]{fig1.eps}
\caption{Insert the figure caption directly after the header. 
Use the words {\it Top, Bottom, Left, } and {\it Right} (in italics) 
to denote the sub figures being described. 
Additional annotation should also be in italics, 
e.g., {\it solid symbols, open symbols, red, dashed line,} etc. If including color figures, please supply a B\&W or greyscale figure as well for print and include the following statement at the end of your caption: (Color figure online.)}
\end{center}
\label{fig1}
\end{figure}

\subsection{Equations}

Place equations in the text and number them, for example:
\begin{equation}
	\frac{\rho-\rho_{0}}{\rho_{0}} = aH + bH^{2},
\end{equation}
When referring to equations please use parentheses and the number of the equation as follows: Eq. (1).

\subsection{Figures}

When inserting figures, Leave an empty line after the Figure caption. When referring to figures in the text, please do so as follows: Fig.~\ref{fig1}.

\section{Section Title}

\subsection{References}
Regarding references: please see the example bibliography below. Please include DOI numbers wherever possible. Please number your references in order of appearance in the text. Citations should be with square brackets as follows [1], [2,3] or [4-7]. 
Please use the commonly used and accepted journal abbreviation whenever possible. For example, the Journal of Low Temperature Physics is J. Low Temp. Phys. \\
When citing articles that shall appear in the same special issue, please do the following. During the review stage you will be asked to cite the article as A. Name, J. Low Temp. Phys. This Special Issue (YEAR). Upon acceptance of the article and receiving the proofs, you will be requested to update the references and replace ''This Special Issue'' with the full citation and DOI number. Articles that have been accepted and published online (on SpringerLink) will immediately have a DOI number even if they are not assigned to an issue yet. When citing articles that appeared in previous special issues of this journal, please use the standard citation and include the DOI number.

\begin{acknowledgements}
(Optional) Please enter text directly after the header. Template v.1 by KLL - June 18, 2015.
\end{acknowledgements}

\pagebreak